**Biomineral Amorphous Lasers through Light-Scattering Surfaces Assembled by Electrospun Fiber Templates**

*Maria Moffa*[1,*], *Andrea Camposeo*[1,*,**], *Vito Fasano*[2], *Barbara Fazio*[3], *Maria Antonia Iatì*[3], *Onofrio M. Maragò*[3], *Rosalba Saija*[4], *Heinz-Christoph Schröder*[5], *Werner E. G. Müller*[5] and *Dario Pisignano*[1,2,**]

[1] NEST, Istituto Nanoscienze-CNR, Piazza S. Silvestro 12, I-56127 Pisa, Italy

[2] Dipartimento di Matematica e Fisica "Ennio De Giorgi", Università del Salento, via Arnesano, I-73100 Lecce, Italy

[3] CNR-IPCF, Istituto Processi Chimico-Fisici, Viale F. Stagno D'Alcontres, 37, I-98158 Messina, Italy

[4] Dipartimento di Scienze Matematiche e Informatiche, Scienze Fisiche e Scienze della Terra, Università di Messina, Viale F. Stagno d'Alcontres 31, I-98166 Messina, Italy

[5] Institute for Physiological Chemistry, University Medical Center of the Johannes Gutenberg University, Duesbergweg 6, D-55099 Mainz, Germany

[*] These authors equally contributed to this work
[**] Corresponding authors: andrea.camposeo@nano.cnr.it, dario.pisignano@unisalento.it

**Abstract**

New materials aim at exploiting the great control of living organisms over molecular architectures and minerals. Optical biomimetics has been widely developed by microengineering, leading to photonic components with order resembling those found in plants and animals. These systems, however, are realized by complicated and adverse processes. Here we show how biomineralization might enable the one-step generation of components for amorphous photonics, in which light is made to travel through disordered scattering systems, and particularly of active devices such as random lasers, by using electrospun fiber templates. The amount of bio-enzymatically produced silica is related to light-scattering capacity and the resulting organosilica surfaces exhibit a transport mean free path for light as low as 3 µm, and lasing with linewidth below 0.2 nm. The resulting, complex optical material is characterized and modelled to elucidate scattered fields and lasing performance. Tightly-controlled nanofabrication of direct biological inspiration establishes a new concept for the additive manufacturing of engineered light-diffusing materials and photonic components, not addressed by existing technologies.

## 1. Introduction

Nature represents an unlimited font of inspiration for the development of novel materials. The exceptional control that living organisms exert over the composition, structure and morphology of many compounds and minerals, the generally mild conditions required to physiologically form these materials, as well as their resulting physical properties may enable smart applications in optics and photonics [1-4]. Two concepts are being developed in this framework. The first involves optical biomimetics through microengineering, namely using or mimicking photonic crystals, anti-reflectors, iridescent structures or other resonators found in many plants and animals [4-6]. The second option relies on the elaboration of manufacturing processes of true biological derivation, namely on the emulation of natural nanofabrication mechanisms, which might involve genetic encoding, specific peptide sequences, or proteins directing the polycondensation of mineral precursors [7-12]. Most often, however, the structures produced *in-vitro* are very far from reaching the richness and the optical functions of natural archetypes.

To achieve truly complex optical materials by biomineralization, it is important to take as much advantage as possible of the characteristic features which are inherent to natural biophotonic structures. These systems are generally *disordered* to some extent, and minerals in them are frequently incorporated in branched microenvironments with low-refractive index ($n \sim 1.5$) compounds, and templated by fibrillar protein architectures in the extracellular matrices. All these properties might address biomineralization in an effective way to realize surfaces and devices for amorphous photonics, exploiting multiply scattered light in disordered structures [13]. For instance, sponges and diatoms are known to synthesize nano-structured silica, by means of special classes of proteins [2,9,10-14]. Silica is widely used in optics, although its artificial production usually involves high temperature and harsh chemical conditions. Among potential applications of silica nanostructures in amorphous photonics, so-called random lasers [6,15,16] have attracted great attention due to the technological potential of these low-cost light sources for medicine and spectroscopy [13], and to the wealth of physical effects occurring when light travels through disordered systems, which include coherent back-scattering (CBS) [16-18] and formation of random resonators in gain materials [16,19].

Here we report on random laser devices realized by biomineralization. Light-diffusing organosilica fibers at the base of the laser formation are obtained by a straightforward method. We developed and compared different surface functionalization strategies of the silicatein protein onto electrospun fibers, followed by *in-vitro* biosilicification. The capacity of building effective light-scattering surfaces is found to be promoted by silica particles at a specific spatial density. The resulting hybrid material incorporated with an organic dye show random lasing spikes with a spectral linewidth smaller than 0.2 nm and a threshold fluence of 7.5 mJ cm$^{-2}$.

The importance of this method stands in various aspects. Firstly, the possibility is opened to direct the realization of functional components for amorphous photonics by cheap and gentle, bioinspired approaches, mimicking the routine assembly of micro-structures that is performed in nature. Secondly, the approach combines the advantages of large-area fiber spinning and biomineralization, thus establishing more versatile design rules for the development of photonic structures. The resulting, silica-fibers hybrid devices exhibit a transport mean free path of light, (3.3 ±0.2) μm, comparable with the smallest values so far reported for low-$n$ materials, which is reflected in a threshold for random lasing smaller than those of other organic solid-state systems.

## 2. Experimental Methods and Materials

### 2.1. Materials

Silicatein-α cDNA from *S. domuncula* is inserted into the oligohistidine expression vector pQ30 (Qiagen), transforming *E. coli* host strain Novagen BL 21 (Merck) with this plasmid and harvesting it in a BIOSTAT Aplus bioreactor (Sartorius Stedim Biotech). Isopropyl β -D-thiogalaktopyranosid is added to induce the expression of the fusion protein, and the recombinant protein is finally collected, purified by affinity chromatography and refolded. Other chemicals are obtained from Sigma Aldrich, unless otherwise specified.

### 2.2. Electrospinning and biosilica synthesis

Solutions with concentration of 200 mg/mL are prepared by dissolving polycaprolactone (PCL, average Mn 80,000) in a dichloromethane:dimethylformamide mixture (80:20 v/v) under stirring for 12 h at room temperature. For carrying out electrospinning the solution is loaded in a 1 mL plastic syringe and injected through a stainless-steel blunt needle using an infusion pump (Harvard Apparatus) at an injection rate of 0.5 mL·h$^{-1}$, with an applied voltage of 5 kV (EL60R0.6-22, Glassman High Voltage). Fibers are collected at a distance of 10 cm from the needle tip, at air humidity of about 45% and temperature of 20°C, and stored under vacuum at room temperature. Fiber functionalization is performed according to the processes schematized in Fig. 1. For methods based on physisorption (*PA*) and chemical grafting (*CG*), fibers are firstly treated with plasma oxygen by a tabletop system (Tucano, Gambetti Kenologia) at 30 W for 3 min. Afterwards, fibers are incubated in a 50 µg/mL silicatein solution for 3 hours. For *CG*, samples are also treated with 10 mM ethyl dimethylaminopropylcarbodiimide hydrochloride (EDC) and 20 mM *N*-hydroxysulfosuccinimide for 1 hour at room temperature prior to protein incubation. For functionalization with polydopamine coatings (*DOPA*), fibers are firstly immersed in 2 mg/mL dopamine 10 mM Tris-HCl (pH 8.5) with gentle shaking for 24 h at room temperature, followed by washing with water to remove unbound dopamine. Samples are then transferred into silicatein solutions (50 µg/mL) for 3h. Pristine PCL fibers are used as negative control. A QuantiPro bicinchoninic acid (BCA) Assay Kit is used to investigate the amount of adsorbed silicatein, measuring absorbance at 562 nm by a spectrophotometer (Lambda 950, Perkin Elmer Inc.). As reference, a 6 points standard curve in triplicate is calculated using silicatein solutions. Each quantification measurement is repeated on at least ten normally identical samples. The concentrations per unit surface area (µg cm$^{-2}$) of protein immobilized onto different surfaces are so calculated. Horizontal attenuated total reflectance-Fourier transform infrared (HATR-FTIR) spectroscopy is carried out by a Spectrum 100 system (Perkin Elmer Inc.), with 2 cm$^{-1}$ resolution and utilizing a ZnSe 45-degree flat-plate. Spectra are baseline-corrected and smoothed. To analyze the protein secondary structure, the second-derivative of spectra are calculated and then fitted with Gaussian band profiles. Fluorescein isothiocyanate (FITC) in dimethyl sulfoxide is used to investigate the silicatein distribution by confocal laser scanning microscopy. To this aim, a bicarbonate buffer (NaHCO$_3$ solution $A$; 3.5 mL, 0.1 M) is mixed with carbonate buffer (Na$_2$CO$_3$ solution $B$; 315 µL, 0.1 M), samples are incubated in $A+B$ (1.3 mL) with addition of FITC/dimethyl sulfoxide solution (100 µL, 1.5 mg/mL), at 4°C for 8 hours. Afterwards, they are washed 3 times with phosphate buffer solution (5 minutes each) and observed by an inverted microscope (Eclipse Ti) equipped by a confocal A1 R MP system (Nikon). Silicatein-bound fibers are incubated in tetraethoxysilanes (TEOS, purity ≥ 99.0%) at room temperature, for times up to 5 days. After 1, 3 and 5 days samples are washed with ethanol and inspected. The morphology and composition of obtained biosilica is evaluated by scanning electron microscopy (SEM) and energy dispersive X-ray (EDX) analysis (FEI). To evaluate the average distance between silica particles, the WSxM software package [20] is used to analyze the SEM images.

### 2.3. Random lasing devices

Biosilicificated fibers are coated with a film of polyvinylpyrrolidone (PVP, Alfa Aesar, Mn 130,000) doped with Disodium-1,3,5,7,8-pentamethylpyrromethene-2,6-disulfonate-difluoroborate complex (pyrromethene, stimulated emission cross-section, $\sigma_{se} \cong 10^{-19}$ cm$^2$, see Supporting Information). The solution is obtained by 30 mg/mL of PVP and 3 mg/mL of pyrromethene in water with sonication for 2 h, and drop-cast onto fibers, up to complete water evaporation at room temperature ($\sim$ 2 days). Light-diffusing surfaces are then excited by the third harmonic of a pulsed Nd:YAG laser ($\lambda_{exc}$ =355 nm, repetition rate=10 Hz, pulse duration=10 ns). The excitation laser beam is focused on the samples in a stripe shape (2 mm $\times$ 150 µm), and precisely aligned on one edge of the samples by means of a micrometric translation stage. The emission signal is collected from the excited edge of the sample by a lens and coupled to an optical fiber. The spectral features of the collected emission are analysed by a monochromator, equipped with a charge coupled device (CCD) detector. The samples are mounted in a vacuum chamber and all measurements are performed in vacuum ambient (pressure $< 10^{-4}$ mbar) in order to minimize degradation effects due to photo-oxidation.

### 2.4. Characterization of light transport properties in the fiber-biosilica system

CBS experiments are performed by a continuous wave He-Ne laser source ($\lambda$=633 nm), a system of lenses, a cube beamsplitter, a $\lambda$/4 waveplate and a polarization filter, and by determining angular profiles of the intensity backscattered from samples by a CCD. For fitting CBS data, we set a slab thickness of 200 µm, and the inelastic mean free path (1700 µm) is calculated for effective media given by either 22%/78% PCL/air or 22%/2%/76%

PCL/silica/air, mimicking actual samples. A finite slab model is used to fit CBS data and determine the transport mean free path for light. Light-scattering in the fiber-biosilica hybrid materials at 355 nm and at 570 nm is studied by field expansion in terms of spherical multipole fields and calculating the T-matrix for clusters of spheres encompassing linear aggregates to describe nanofibers.

### 3. Results and Discussion

#### 3.1. Biomineralized surfaces on electrospun fiber templates

A platform of different techniques is implemented in this work to functionalize electrospun PCL fibers with silicatein (Fig. 1), including *PA*, *CG* and *DOPA* methods. Aspecific, *PA*-functionalization is performed by direct incubation in the protein solution (Fig. 1a), whereas EDC and sulfo-*N*-hydroxysulfosuccinimide are used as zero-length crosslinker and catalyst for *CG*, respectively (Fig. 1b). Such coupling procedure immobilizes proteins onto supports without introducing spacing arms. Finally, polydopamine coatings on PCL are realized to subsequently immobilize silicatein by simple dipping (Fig. 1c). This method, based on the oxidative polymerization of dopamine, lacks of complicated process steps and of possible cleavage of polymer chains, thus potentially overcoming a few important drawbacks of common surface modifiers [21].

The amount of protein adsorbed on the electrospun templates is then estimated by labeling fiber-bound silicatein with FITC for confocal laser scanning microscopy, evidencing a more uniform distribution and a higher amount of proteins along fibers undergone *CG* and *DOPA*-treatments (Figure 2a-h and Figure S1 in the Supporting Information). These results are supported by quantification by a BCA assay (Fig. 2i). *CG* and *DOPA*-methods allow significantly higher amount of immobilized protein to be found ($22\text{-}23\pm2$ μg/cm$^2$) compared to *PA* ($5\pm1$ μg/cm$^2$). However, we point out that the bare amount of immobilized protein does not necessarily indicate what is the best functionalization method in terms of leading to effectively light-scattering surfaces. In particular, a higher amount of immobilized proteins does not necessarily result in better-suited biosilicificated mats for random lasing, as will be assessed in the following.

To investigate the surface-silicatein interactions more in depth, we study the different samples by HATR-FTIR spectroscopy (Figure S2). Characteristic infrared bands for PCL-related stretching modes are notable for pristine PCL, polydopamine-treated PCL (PCL*, without silicatein), and silicatein-functionalized fibers (*PA*, *CG* and *DOPA* samples), with the latter also showing peaks at 3300 cm$^{-1}$ (1 in Fig. S2), 1650 cm$^{-1}$ (2) and 1560 cm$^{-1}$ (3), attributable to amide A, amide I and amide II, respectively. The silicatein secondary structure can be analyzed by the shape of the amide I band. To this aim, the relative areas of the overlapping band components under the amide I contour are resolved and quantified through their second-derivative (Fig. 2j), in combination with Gaussian curve-fitting analysis [22]. Bands at 1693, 1639, and 1628 cm$^{-1}$ are to be assigned to β-sheets, the band at 1652 cm$^{-1}$ to an α-helix, bands at 1684, 1671, and 1663 cm$^{-1}$ to turns, and the band at 1645 cm$^{-1}$ to an unordered structure, respectively [23]. The quantification of these bands (Fig. 2k) evidences for polydopamine and *PA*-methods higher levels of secondary structures, associated to the presence of active silicateins. These features, related to the different reactivity of fibers toward proteins and to consequent conformational changes of silicateins, are likely to affect the resulting bioenzymatic activity. Taken together, these results highlight polydopamine-mediated functionalization as optimal approach ensuring both high amounts of proteins linked to electrospun templates and enhanced bands associated to pristine secondary structures.

Spun templates, including uncoated samples as control, are then incubated with TEOS at ambient conditions. The resulting formation of biosilica on protein-functionalized samples is directly observable by SEM performed at different time steps of incubation (Figure 3). No deposits are appreciable on pristine PCL fibers even after five days (Fig. 3a, e and i). Instead, on silicatein-functionalized fibers, silica spheres with diameter up to 300 nm start to precipitate during the first day of TEOS incubation (Fig. 3b-d). The density of spheres found on *PA*-samples after 1 day and 5 days is of about $1.5 \times 10^6$ particles/mm$^2$ and $4 \times 10^6$ particles/mm$^2$, respectively. A relatively higher density is observed on *CG* and *DOPA*-samples (i.e. $\sim 3.5 \times 10^6$/mm$^2$ and $6 \times 10^6$/mm$^2$ after 1 day). Upon increasing the incubation time, the area covered by biosilica increases as well, up to completely covering *CG* and *DOPA*-samples with a continuous film (Fig. 3k,l). The different amount of deposited silica will have a strong impact on the resulting lasing properties of these surfaces, as described below. In addition, EDX analysis of the synthetized material gives evidence that it consists of silica, whereas no silica is formed on pristine fibers incubated in the absence of silicatein (Fig. 3m-p). Finally, silica particles are quite uniform. For instance, those formed on *PA*-samples are regular spheres with average diameter about 180 nm. These results are in agreement with previous studies on histidin-tagged silicatein catalyzing the formation of interconnected silica nanospheres with a diameter up to 300 nm [24], and are well consistent with *in vivo* biosilicification in sponges, in which spicules formation is initiated by the synthesis of silica granules with size of 100-600 nm.

### 3.2. Random lasers

Surfaces embedding wavelength-scale spheres can be excellent diffusing elements leading to random lasers, depending on the degree of surface coverage. Figure 4 and Figure S3 summarize the emission properties of the various biosilicificated surfaces under ns-pulsed excitation, following the deposition of a film of PVP doped with pyrromethene. In particular, in Fig. 4 we compare (i) devices where only electrospun (PCL) fibers are used in addition to the dye-doped polymer supporting optical gain, with (ii) devices where fibers are coated with biomineralized silica spheres prior dye-doped polymer casting. The two devices are therefore identical in their fibrous geometry underneath and gain material, namely their eventually different spectral and threshold behavior is directly related to the presence of additional scattering from silica spheres, at a well-defined surface density, in biomineralized samples. Therefore, this comparison is highly useful in view of elucidating the working mechanisms of the random laser. Fig. 4a and d, and Fig S3a and d display the intensity maps showing emission spectra as a function of the excitation fluence, for pristine PCL fibers, *PA*, *CG* and *DOPA* samples (photographs are in the corresponding insets, and highlight a quite different morphology even at macroscale). What is found is that in pristine PCL, as well as in *CG* and *DOPA* samples, emission line-narrowing occurs upon increasing the excitation fluence above roughly 10, 23 and 34 mJ cm$^{-2}$, respectively, with a smooth transition across a range of 5-10 mJ cm$^{-2}$ as typical of amplified spontaneous emission (ASE, Fig. 4b and Fig. S3b,d). In these samples the emission intensity ($I$) has a nonlinear, threshold-less dependence on the excitation density, which is well described by an exponential-like relation as expected for ASE. Overall, these systems exhibit good optical gain (up to tens of cm$^{-1}$). However, even *CG* and *DOPA* devices lack of the effective feedback mechanism needed for lasing, due to the formation of the continuous and dense layer of silica spheres. Such layer favors the waveguiding of the spontaneous emission along the organic layer, and the occurrence of ASE-related gain narrowing instead of random lasing.

*PA*-samples, which features a less dense population of silica microspheres compared to the other functionalized surfaces, show a different behaviour. Upon increasing the excitation fluence, a peak centred at about 569 nm and with linewidth of a few nm features a threshold behaviour (Fig. 4e,f). Such threshold (7.5 mJ cm$^{-2}$) can be easily determined by considering the linear dependence of the emission intensity on the excitation fluence above threshold. In addition, a sharp line-narrowing transition is found at threshold. Under single-shot excitation, the laser shows a complex dynamics, with the spectral position of appreciable narrow peaks whose spectral position may vary at each pumping pulse [15] (Figure 5a). Additional spikes are also visible on the top of the smooth peak in Fig. 4a, and show a sharp linewidth reduction down to 0.2 nm (full width at half maximum) at threshold (Fig. 5b), as estimated by subtracting the underlying broader spectral background.

The comparison of excitation threshold performance shown in different studies on random lasers is made very hard by its dependence on experimental parameters such as the area of the excited region on samples. However, here found threshold values are lower than those of other random lasers based on low-$n$, organic systems, such as biopolymers embedding non-linear chromophores [25], and are in line with recently reported, plasmon-enhanced hybrid architectures, such as dye-doped fibers incorporating Au nanoparticles [26], and polymer films on three-dimensional nanorod metamaterials [27]. The working mechanisms of random lasing supported by the biomineralized surface involve two different effects as schematized in the inset of Fig. 5b, namely stimulated emission from the active material as well as scattering from light-diffusing components. The active material is located in the dye-doped PVP layer and, in absence of significant scattering of generated light, would produce ASE supported by waveguiding along this organic slab as commonly found in light-emitting organic semiconductor films [28, 29]. The thickness of the active layer is an important parameter, which when above a cut-off value ($t_C$) supports modes at a given wavelength ($\lambda$), $t_C=\lambda\times\{\arctan [(ns^2-1)/(n^2-ns^2)]^{1/2}\}/[2\pi(n^2-ns^2)^{1/2}]$, where $n$ and $ns$ are the refractive indexes of the dye-doped PVP and of the biomineralized material, respectively. For our system, $t_C \cong 500$ nm. When biomineralization occurs and silica particles are achieved at a given density, additional effects take place in the device. On one side, electrospun fibers can also guide coupled emitted light along their londitudinal axis [30], namely along random directions. More importantly, significant light-scattering might take place, in principle due to both electrospun fibers and silica particles. The contribution from individual fibers in diffusing light far from the pristine propagation direction can be estimated to be small. Indeed, the calculation of the angular dependence of the light-scattering form factor for polymer fibers [31], highlights that most of the incident light is scattered at forward angles (≤40°) when, as in our case, $ka$ is comparable or larger than unity, where $a$ is the fiber radius and $k=2\pi/\lambda$ is the wavevector. Electrospun PCL filaments therefore do not substantially diffuse emitted light at large angles, which is also supported by the ASE-like emission features found for the device realized on pristine PCL fibers (Fig. 4b,c). Biosilica particles, instead, more significantly affect the transport of light and hence the overall emission. The average distance between first-neighbour particles is especially relevant in this respect. When such distance is at least 280 nm, as in $PA$-samples, a well-defined threshold is found to exist for random lasing (Fig. 4e). Instead, line-narrowing is observed at excitation fluences higher by 3 to 5 times (up to about 45 mJ cm$^{-2}$, Fig. S3b,e) when the inter-particle separation becomes lower than 250 nm and ultimately leads to densely-packed biosilica spheres, arranged in a continuous layer, as in $CG$ and $DOPA$-samples. In a whole, these findings indicate a quite weak effect of light-scattering from fibers at the lasing wavelength, with a much more relevant role played by scattering from silica particles, and also suggest that the interface between the active film and the scattering surface is highly important to provide the refractive index contrast and the enhanced fields needed for lasing.

The light propagation in the hybrid material is clearly non-ballistic [32], the excited sample size (> $10^2$ μm) being much larger than the average distance travelled by photons between consecutive scattering centers (inset of Fig. 5b). This leads to an amplification length, $l_A$, in the device, much shorter than the gain length ($l_G$) in the active material, since multiple scattering might shorten the distance along the propagation direction over which the light is amplified in an effective way. In our system, $l_G$ is of the order of a few hundreds of μm as found by ASE measurements (Supporting Information), whereas $l_A$ is roughly given by $(l_G \, l_T)^{1/2}$ and can also be estimated from the transport mean free path, $l_T$.

### 3.3. Light-scattering properties

To better assess light transport properties and estimate the transport mean free path provided by the biomineralized material, the random lasing, hybrid *PA*-system is studied both experimentally and theoretically and compared to those of pristine networks of PCL fibers. Firstly, the intensity of back-scattered light from 200 μm thick samples is measured by irradiating the diffusing material by a He-Ne laser. We then analyze the CBS cones, which describe the ratio of the total scattered light intensity to the diffuse background as a function of the angle, θ, defined by the incoming and the outgoing light wavevectors. An enhanced reflectance is expected at small θ values following constructive interference of light along time-reversed optical paths and related to weak localization of light [17]. The here measured enhancement factor ($A_{CBS}$) at the exact backscattering direction due to interference in the complex material is lower than the theoretical value of 2 due to possible residues of single scattering events and stray light. Furthermore the slight decrease of $A_{CBS}$ following silica deposition could suggest the incipient occurrence of recurrent multiple scattering events, folded and loop, in the stronger scattering material. Experimental CBS data are shown in Figure 6a, and evidence a larger CBS cone, namely a shorter transport mean free path for light (larger scattering strength), in fibers with silica particles. Data are well fitted by the function, $I(\theta) = [1 + (A_{CBS} - 1)(\gamma_C / \gamma_L)]$ (Supporting Information), where $\gamma_C$ and $\gamma_L$ are the θ-dependent coherent and the diffuse background intensities for a finite slab, respectively, expressed in terms of the measured fluxes of photons per solid angle and per probed area normalized to the incident fluxes (so-called bistatic coefficients) [33]. The silica-fibers hybrid structure exhibits a transport mean free path of light as low as (3.3 ±0.2) μm, comparable with the smallest values for low-$n$ materials found in dense fibrillar chitin [34]. This would lead to a $l_A$ value of about 30 μm, which indicates the effectiveness of the amplification mechanism promoted by the light-scattering surface.

These data are indicative of specific light transport properties in the hybrid structure composed of organic fibrils and silica particles. Such complex material encompassing highly elongated filaments is hard to depict as made by isotropic scatterers. Therefore we develop a model for the scattering of light by the transition matrix (T-matrix) technique [35], describing fibers as randomly-directed linear aggregates of 430 nm spherical building blocks (Fig. 6b), and adding silica particles with 180 nm diameter (Fig. 6c) as in *PA*-samples (details in Supporting Information). Biomineralization is very advangeous to this aim, since this method allows for overcoming most of the issues of etched and ground materials in terms of poorly defined shapes of scattering elements [15] and to achieve well-defined single-particle scattering properties [17]. Furthermore, being particles directly formed on the fiber template, uncontrolled clustering due to van der Waals or electrostatic forces in solution is avoided, which also simplifies modelling. Hence, we calculate the maps of scattered field normalized to an unpolarized incident field intensity ($|E_S/E_0|^2$), at both the emission (570 nm, Fig. 6d,e), and the excitation (355 nm, Fig. 6f,g) wavelength. The hybridization with silica nanoparticles is clearly observed to lead to strong scattering, which can be accounted by a better match between the silica particle size and the incident wavelengths [35]. At the same time, organic filaments support fields at the emission wavelength (Fig. 6d) while having a much weaker effect on the excitation which is almost uniformly spread over the sample plane (Fig. 6f). The resulting hot spots, namely enhanced local fields in the *PA*-material suggest more effective excitation of chromophores (Fig. 6g) as well as higher overall feedback which is directly related to the occurrence of lasing at about 570 nm (Fig. 6e). Analogous results are found for any direction of incoming light at the emission wavelength across the material (Figure S4). Finally, the system is effective in distributing scattered pumping fields about 1 µm deep into the sample, thus providing ample space for exciting surrounding dyes (Figure S5). This result supports the importance of the interface between the active material and the scattering surface, which directly involves only the top layers of the fibrous template, and the biomineral component which is there deposited. In this respect, having separated light-scattering and emissive components in our lasers is important in perspective to reach high control on field distribution and lasing modes [36].

### 4. Conclusions

Results presented here indicate that a simple, straightforward method can be developed to realize *in-vitro* nanostructured organo-mineral photonic materials by using electrospun fiber template biomineralization. The approach demonstrates how biosilification can direct the formation of components for amorphous photonics, a route never explored before which leads to the controlled generation of active optical devices such as random lasers. The transport of light in the material show a combined effect of organic filaments and biosilica particles in supporting local field enhancement and lasing. Mimicking natural microstructures by *in-vitro* fabrication promise to greatly simplify design and manufacturing concepts for engineered light-diffusing materials.

*Acknowledgements*
The research leading to these results has received funding from the European Research Council under the European Union's Seventh Framework Programme (FP/2007-2013)/ERC Grant Agreement n. 306357 ("NANO-JETS"). The support from the Apulia Regional Projects 'Networks of Public Research Laboratories' Wafitech (9) and M.I.T.T. (13) is also acknowledged. W.E.G.M. is also holder of an ERC Advanced Investigator Grant (no. 268476 "BIOSILICA").

## References

[1] Y.-F. Huang, S. Chattopadhyay, Y.-J. Jen, C.-Y. Peng, T.-A. Liu, Y.-K. Hsu, C.-L. Pan, H.-C. Lo, C.-H. Hsu, Y.-H. Chang, C.-S. Lee, K.-H. Chen, and L.-C. Chen, Nat. Nanotechnol. **2**, 770 (2007).

[2] K. Yu, T. Fan, S. Lou, and D. Zhang, Prog. Mater. Sci. **70**, 1 (2015).

[3] U.G.K. Wegst, H. Bai, E. Saiz, A.P. Tomsia, and R.O. Ritchie, Nat. Mater. **14**, 23 (2015).

[4] A.R. Parker, and H.E. Townley, Nat. Nanotechnol. **2**, 347 (2007).

[5] K. Liu, and L. Jiang, Nano Today **6**, 155 (2011).

[6] H. Noh, J.K. Yang, S.F. Liew, M.J. Rooks, G.S. Solomon, and H. Cao, Phys. Rev. Lett. **106**, 183901 (2011).

[7] C. Jeffryes, T. Gutu, J. Jiao, and G.L. Rorrer, ACS Nano **2**, 2103 (2008).

[8] J.M. Galloway, J.P. Bramble, and S.S. Staniland, Chem. Eur. J. **19**, 8710 (2013).

[9] K. Shimizu, J. Cha, G.D. Stucky, and D.E. Morse, Proc. Natl. Acad. Sci. USA **26**, 6234 (1998).

[10] N. Kröger, S. Lorenz, E. Brunner, and M. Sumper, Science **298**, 584 (2002).

[11] J.N. Cha, K. Shimizu, Y. Zhou, S.C. Christiansen, B.F. Chmelka, G.D. Stucky, and D.E. Morse, Proc. Natl. Acad. Sci. USA **96**, 361 (1999).

[12] A. Scheffel, N. Poulsen, S. Shian, and N. Kröger, Proc. Natl. Acad. Sci. USA **22**, 3175 (2011).

[13] D. Wiersma, Nat. Photon. **7**, 188 (2013).

[14] M.N. Tahir, P. Théato, W.E.G. Müller, H.C. Schröder, A. Borejko, S. Faiß, A. Janshoff, J. Huth, and W. Tremel, Chem. Commun. **44**, 5533 (2005).

[15] D. Wiersma, Nat. Phys. **4**, 359 (2008).

[16] A. Tulek, R.C. Polson, and Z.V. Vardeny, Nat. Phys. **6**, 303 (2010).

[17] M.P. van Albada, and A. Lagendijk, Phys. Rev. Lett. **55**, 2692 (1985).

[18] P.E. Wolf, and G. Maret, Phys. Rev. Lett. **55**, 2696 (1985).

[19] V.M. Alpakov, M.U. Raikh, and B. Shapiro, Phys. Rev. Lett. **89**, 016802 (2002).

[20] I. Horcas, R. Fernández, J. M. Gómez-Rodríguez, J. Colchero, J. Gómez-Herrero, and A. M. Baro, Rev. Sci. Instrum. **78**, 013705 (2007).

[21] Y.B. Lee, Y.M. Shin, J.H. Lee, I. Jun, J.K. Kang, J.C. Park, and H. Shin, Biomaterials **33**, 8343 (2012).

[22] A. Dong, B. Kendrick, L. Kreilgârd, J. Matsuura, M.C. Manning, and J.F. Carpenter, Arch. Biochem. Biophys. **347**, 213 (1997).

[23] A. Dong, P. Huang, and W.S. Caughey, Biochemistry **29**, 3303 (1990).

[24] M.N. Tahir, P. Théato, W.E.G. Müller, H.C. Schröder, A. Janshoff, J. Zhang, J. Huth, and W. Tremel, Chem. Commun. **24**, 2848 (2004).

[25] L. Sznitko, J. Mysliwiec, P. Karpinski, K. Palewska, K. Parafiniuk, S. Bartkiewicz, I. Rau, F. Kajzar, and A. Miniewicz, Appl. Phys. Lett. **99**, 031107 (2011).

[26] R. Zhang, S. Knitter, S. F. Liew, F. G. Omenetto, B. M. Reinhard, H. Cao, and L. Dal Negro, Appl. Phys. Lett. **108**, 011103 (2016).

[27] Z. Wang, X. Meng, S. H. Choi, S. Knitter, Y. L. Kim, H. Cao, V. M. Shalaev, and A. Boltasseva, Nano Lett. **16**, 2471 (2016).

[28] F. Hide, M. A. Diaz-Garcia, B. J. Schwartz, M. R. Andersson, and Q. B. Pei, Science **273**, 1835 (1996).

[29] A. K. Sheridan, G. A. Turnbull, A. N. Safonov, and I. D. W. Samuel, Phys. Rev. B **62**, 929 (2000).

[30] V. Fasano, A. Polini, G. Morello, M. Moffa, A. Camposeo, and D. Pisignano, Macromolecules **46**, 5935 (2013).

[31] L. Persano, M. Moffa, V. Fasano, M. Montinaro, G. Morello, V. Resta, D. Spadaro, P.G. Gucciardi, O.M. Maragò, A. Camposeo, and D. Pisignano, Proc. SPIE **9745**, 97450R (2016).

[32] H. Cao, Waves Random Media **13**, R1 (2003).

[33] M.B. Van der Mark, M.P. van Albada, and A. Lagendijk, Phys. Rev. B 37, 3575 (1988).

[34] M. Burresi, L. Cortese, L. Pattelli, M. Kolle, P. Vukusic, D.S. Wiersma, U. Steiner, and S. Vignolini, Sci. Rep. **4**, 6075 (2014).

[35] F. Borghese, P. Denti, and R. Saija, Scattering from model nonspherical particles, 2nd Ed., Springer-Verlag Berlin Heidelberg (2007).

[36] M. Leonetti, C. Conti, and C. López, Phys. Rev. A **85**, 043841 (2012).

**Figures**

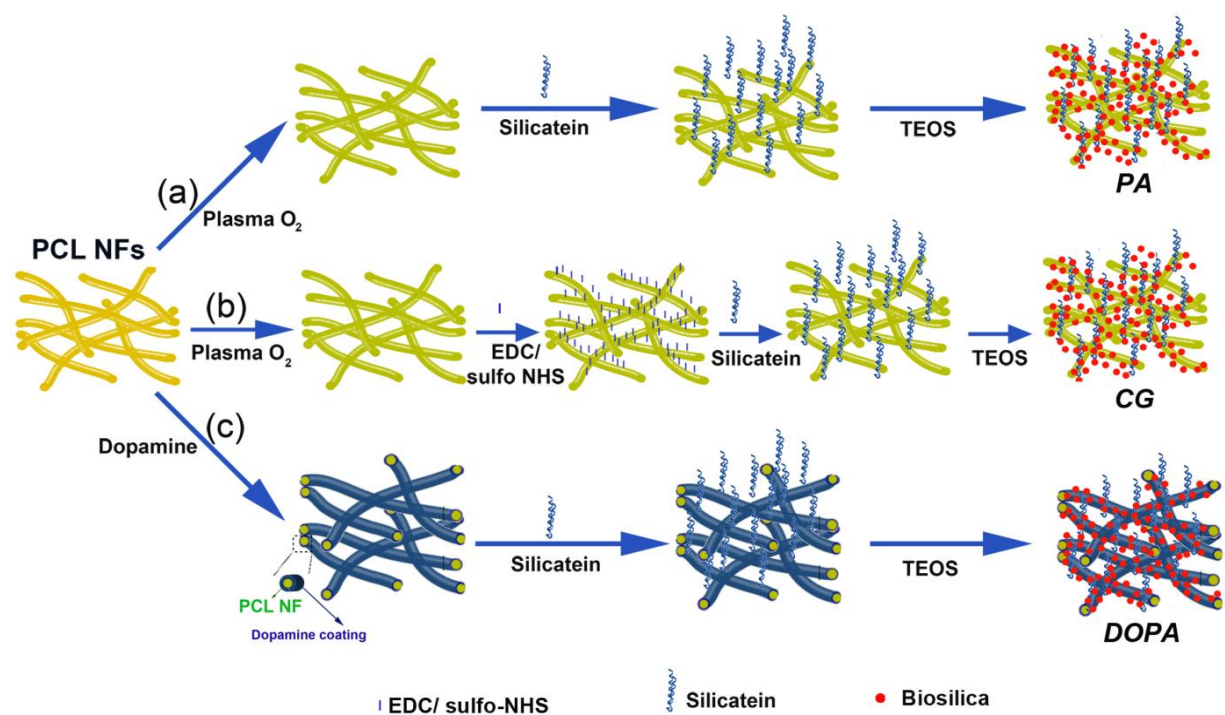

**Figure 1.** Schematics of the stepwise silicatein-coating of nanofibers and production of biosilica over large-area templates. (a) *PA*, (b) *CG* and (c) *DOPA* functionalization.

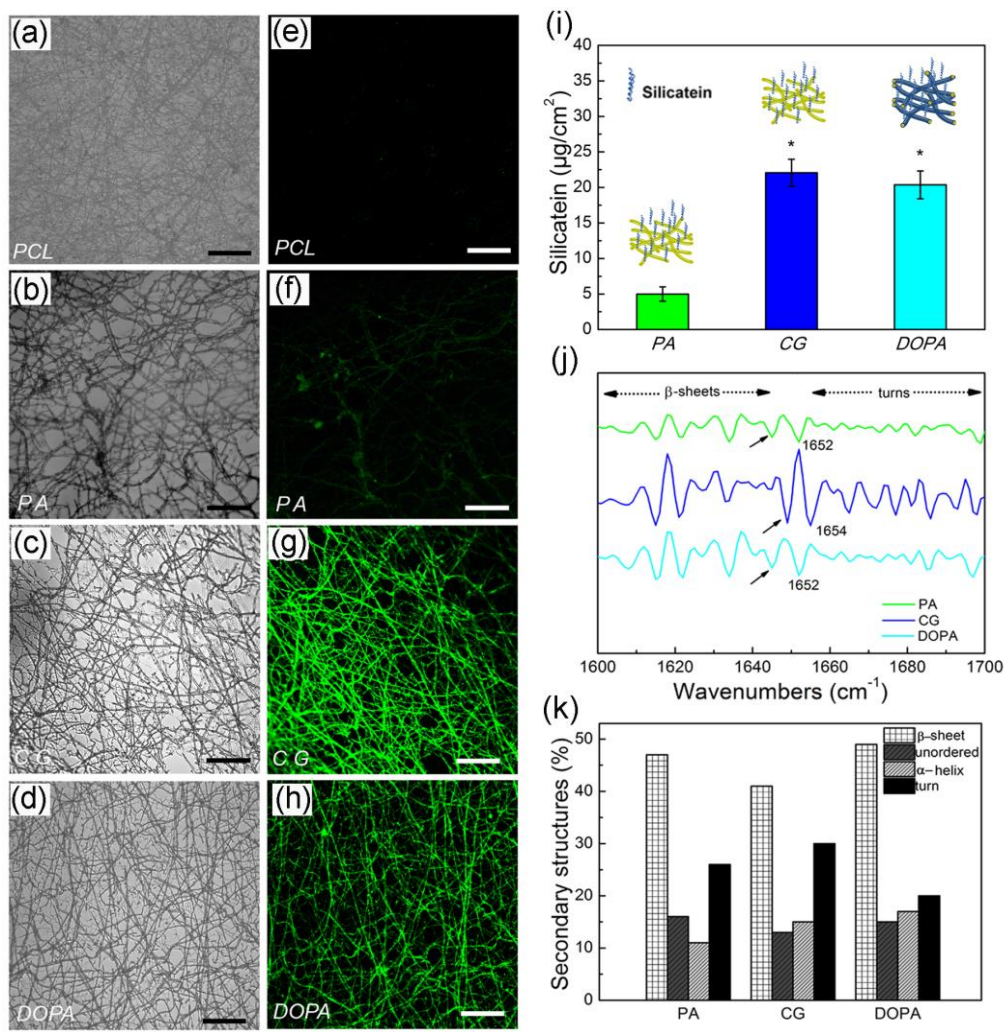

**Figure 2. Recombinant silicatein on electrospun templates.** Optical transmission (a-d) and fluorescence (e-h) micrographs of FITC-labeled silicatein, immobilized on PCL fibers by physisorption (*PA*, b, f), chemical grafting (*CG*, c, g) and polydopamine (*DOPA*, d, h). (a) and (e): pristine PCL fibers treated with FITC, used as control. (i) Quantification of immobilized silicatein on electrospun templates by the different functionalization methods. Results are expressed as (mean ± standard deviation). Asterisks show statistically significant differences compared to *PA* ($P < 0.05$). The cartoons above each column show the different samples. Vertical helices and fiber sheaths schematize silicatein and the dopamine coating, respectively. (j, k) Secondary structure analysis of silicatein adsorbed on nanofibers. (j) Second derivative FTIR spectra of the amide I region, for silicatein-functionalized fibers by physisorption (*PA*, top, green curve), chemical grafting (*CG*, middle, blue curve) and polydopamine (*DOPA*, bottom, cyan curve). (k) Corresponding percent quantification of β-sheet, unordered, α-helix and turns secondary structures of silicatein.

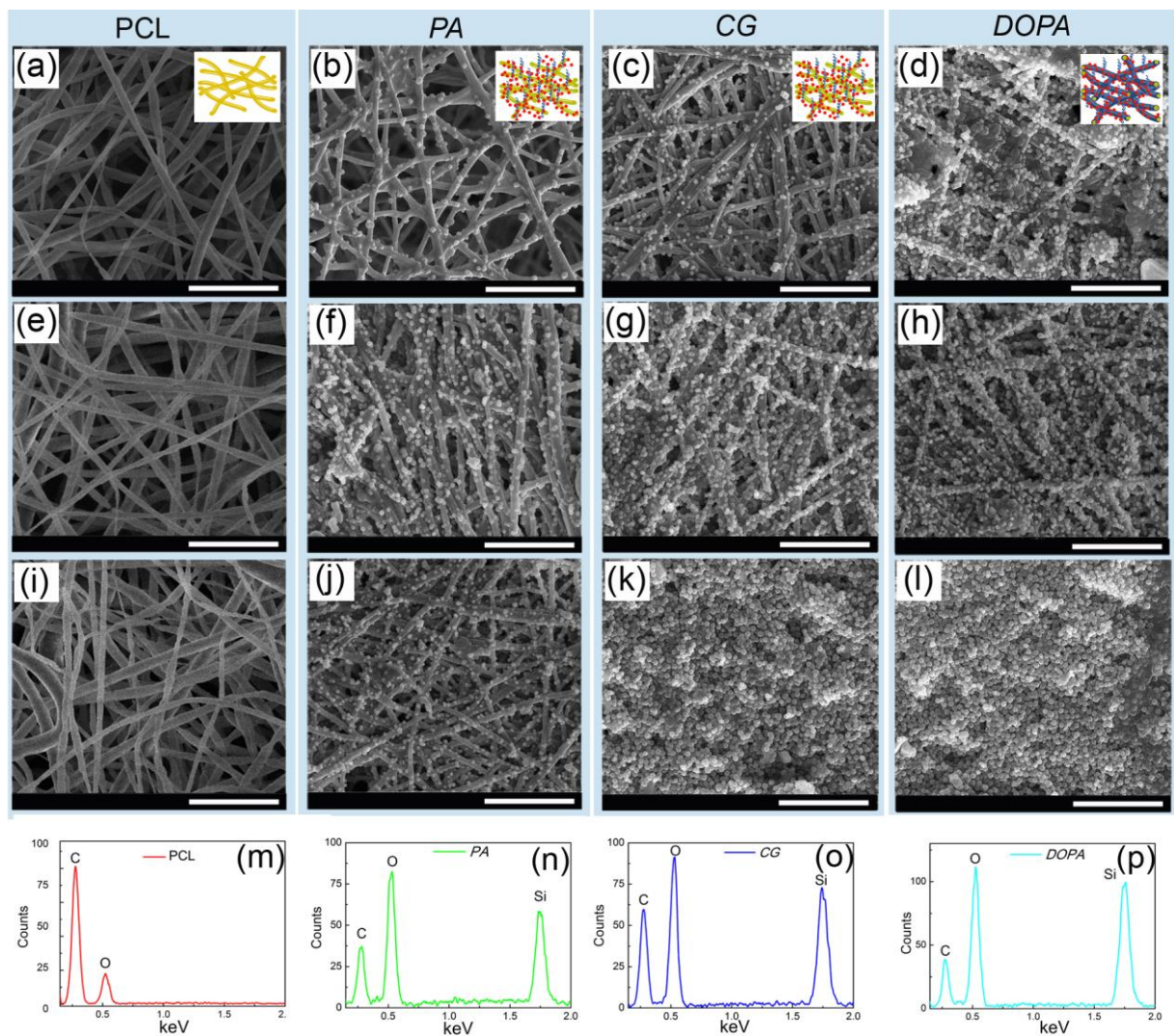

**Figure 3. Synthesis of biosilica spheres on silicatein-functionalized spun fibers.** SEM micrographs of samples obtained by *PA* (b, f and j), *CG* (c, g and k) and polydopamine (*DOPA*, d, h and l), after various days of incubation in TEOS. (a), (e) and (i) micrographs show pristine PCL fibers as control. Fibers are incubated for 1 day (a-d), 3 days (e-h) and 5 days (i-l), respectively. Scale bar = 5 μm. (m-p) Corresponding EDX spectra after 5 days of incubation, for PCL fibers (m), and for samples undergone *PA* (n), *CG* (o) and polydopamine treatment (p). The top-right insets in each column show sample schematics. Vertical helices: silicatein. Fiber sheaths: dopamine coating. Red dots: biosilica.

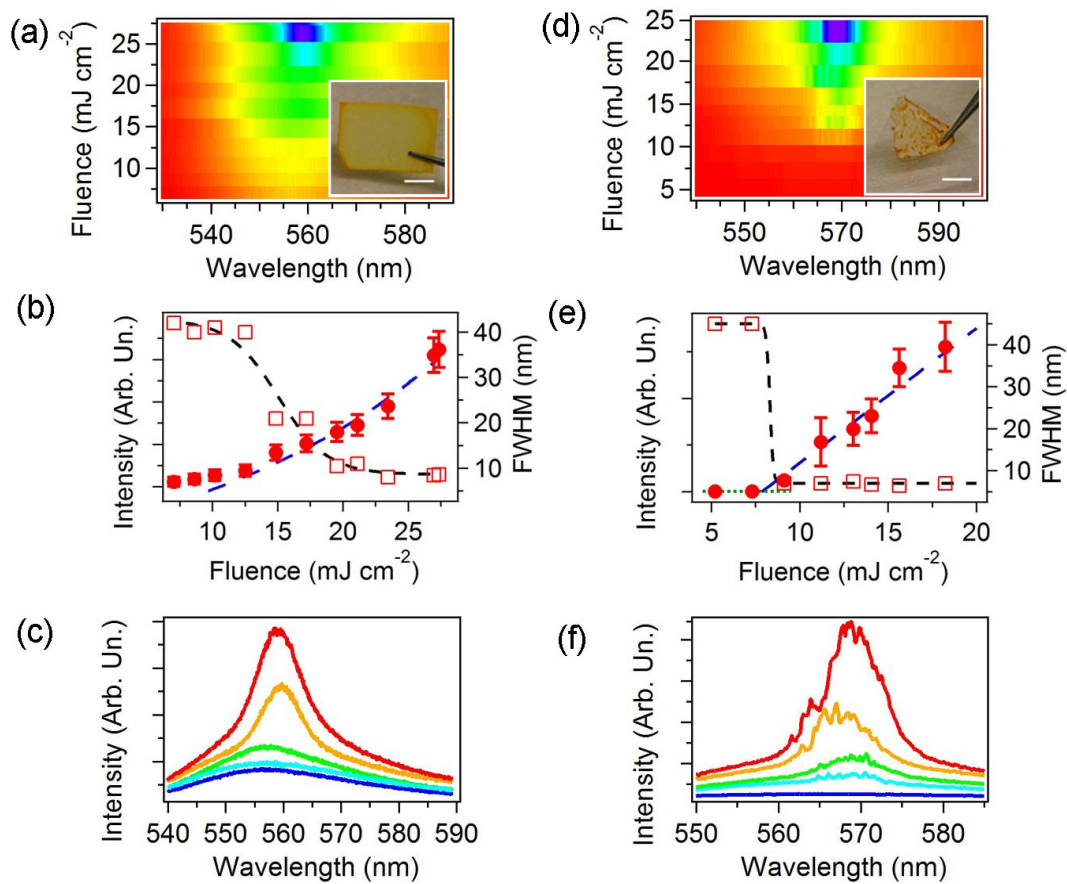

**Figure 4**. **Random lasing of organosilica surfaces**. Analysis of the emission properties, under pulsed excitation, of devices realized on PCL fibers (a-c), and *PA* dye-doped devices (d-f), respectively. (a), (d): Emission intensity maps showing the spectra of the samples as a function of the excitation fluence. Insets: corresponding device photographs. Scale bars = 5 mm. (b), (e): Intensity (left vertical axis) and spectral line-width (right axis) of the emission *vs*. excitation fluence. The blue dashed lines in (b) and (e) are fits to intensity data by an exponential-like law and by an above-threshold linear curve, respectively. The black dashed lines are guide to the eye for line-narrowing data. (c), (f): Emission spectra collected at increasing excitation fluences. Shown curves are averages of many single-shot emission spectra. From bottom to top, excitation fluences: (c): 8, 12, 14, 23, 27 mJ cm$^{-2}$; (g): 5, 11, 14, 18, 23 mJ cm$^{-2}$.

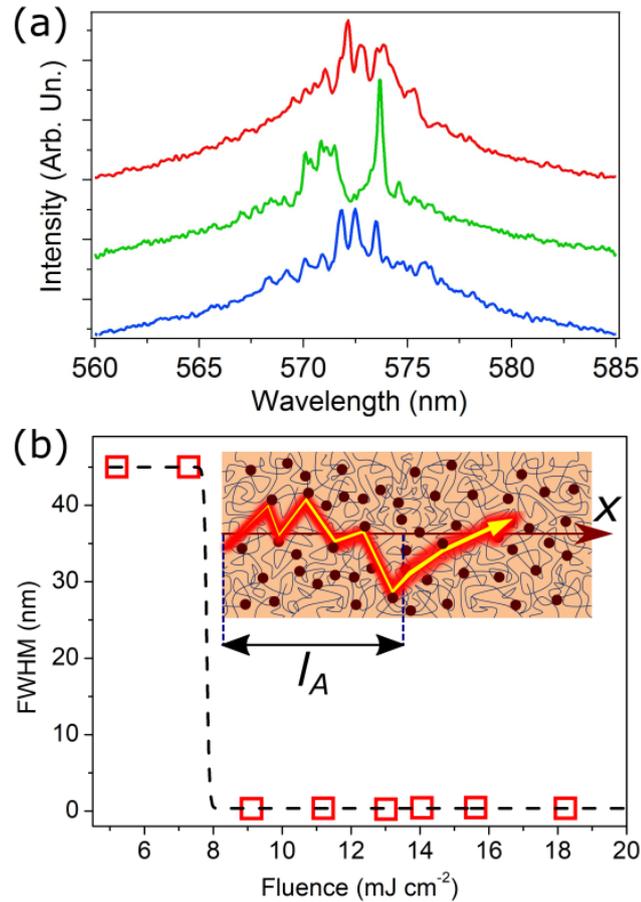

**Figure 5.** (a) Examples of single-shot random lasing spectra from *PA* dye-doped materials. Spectra are shifted vertically for better clarity. (b) Line-narrowing of random lasing spikes on the top of the intensity-feedback mode. The black dashed line is a guide to the eye. Inset: Schematics of the working device, through non ballistic, diffusive regime along the hybrid material. The coloured background stands for the active material. Dark wires: electrospun template fibers. Dark dots: biosilica particles. $x$: main propagation axis for stimulated emission, provided by the length of the excitation stripe. $l_A$: amplification length in the disordered medium, corresponding to an enhancement of the optical signal by a factor $e$. $l_A \cong$ 30 μm in our system.

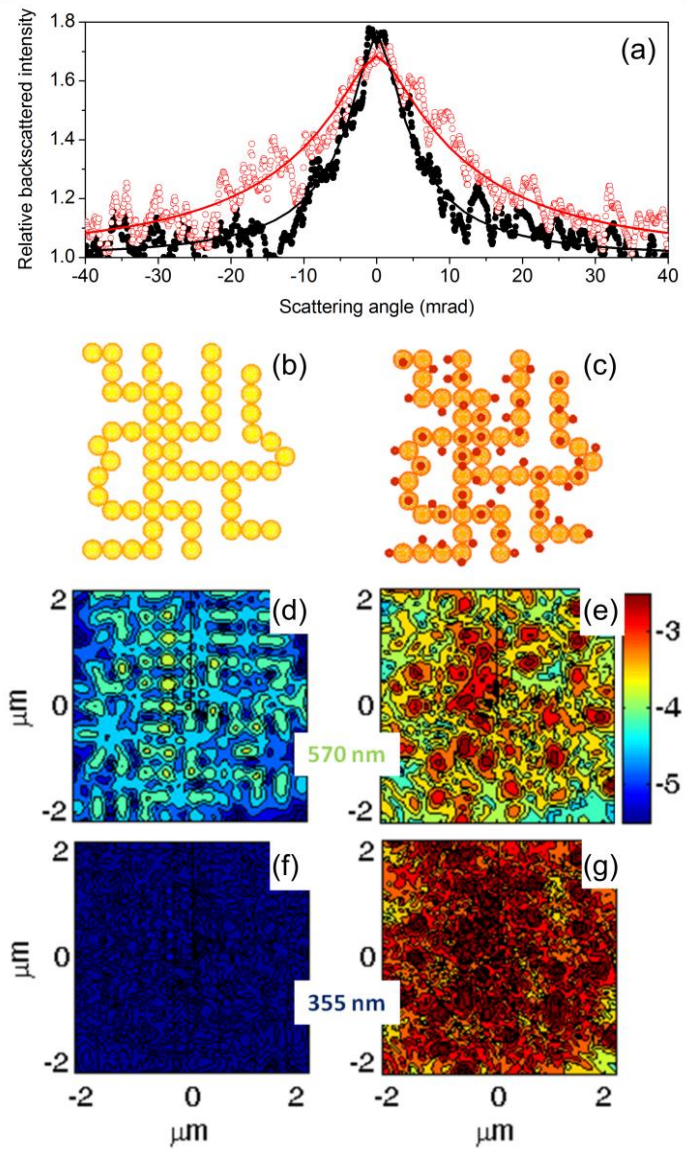

**Figure 6. Light transport in organosilica material.** (a): Experimental CBS data, for pristine PCL fibers (black full dots) and for *PA*-samples (red open dots), and corresponding fits (see the text). Fibers: $A_{CBS}$ = 1.77, transport mean free path = (8.3 ±0.3) μm. *PA*-samples: $A_{CBS}$ = 1.68, transport mean free path = (3.3 ±0.2) μm. Inelastic mean free path for fitting curves: 1700 μm. The absorption of silica is negligible and it does not affect the inelastic mean free path. (b, c): Models adopted for light scattering computation for the nanofiber mats (b) and for the hybridized nanofiber-biosilica structures (c). Filaments and particles are embedded in PVP as external medium. (d-g): Normalized intensity maps, $\left|E_S/E_0\right|^2$, of the scattered field for the fiber mats (d) and for the hybridized structure (e) at the emission wavelength, 570 nm (logarithmic scale). Normalized intensity maps of the scattered field for the fiber mats (f) and for the hybridized structure (g) at the excitation wavelength, 355 nm.

# Supporting Information

**Biomineral Amorphous Lasers through Light-Scattering Surfaces Assembled by Electrospun Fiber Templates**

*Maria Moffa[1,\*], Andrea Camposeo[1,\*,\*\*], Vito Fasano[2], Barbara Fazio[3], Maria Antonia Iatì[3], Onofrio M. Maragò[3], Rosalba Saija[4], Heinz-Christoph Schröder[5], Werner E. G. Müller[5] and Dario Pisignano[1,2,\*\*]*

[1] NEST, Istituto Nanoscienze-CNR, Piazza S. Silvestro 12, I-56127 Pisa, Italy
[2] Dipartimento di Matematica e Fisica "Ennio De Giorgi", Università del Salento, via Arnesano, I-73100 Lecce, Italy
[3] CNR-IPCF, Istituto Processi Chimico-Fisici, Viale F. Stagno D'Alcontres, 37, I-98158 Messina, Italy
[4] Dipartimento di Scienze Matematiche e Informatiche, Scienze Fisiche e Scienze della Terra, Università di Messina, Viale F. Stagno d'Alcontres 31, I-98166 Messina, Italy
[5] Institute for Physiological Chemistry, University Medical Center of the Johannes Gutenberg University, Duesbergweg 6, D-55099 Mainz, Germany
[\*] These authors equally contributed to this work
[\*\*] Corresponding authors: andrea.camposeo@nano.cnr.it, dario.pisignano@unisalento.it

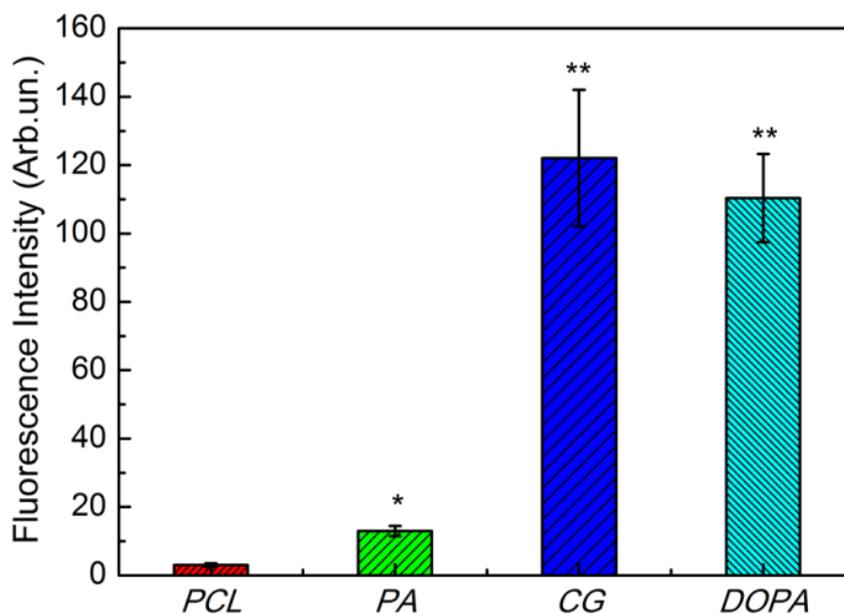

**Figure S1**. Comparison of the fluorescence intensity from FITC-labeled silicatein on PCL nanofiber templates processed by different functionalization methods: physisorption (*PA*), chemical grafting (*CG*) and polydopamine coating (*DOPA*). * show statistically significant differences (*$P < 0.05$; **$P < 0.01$).

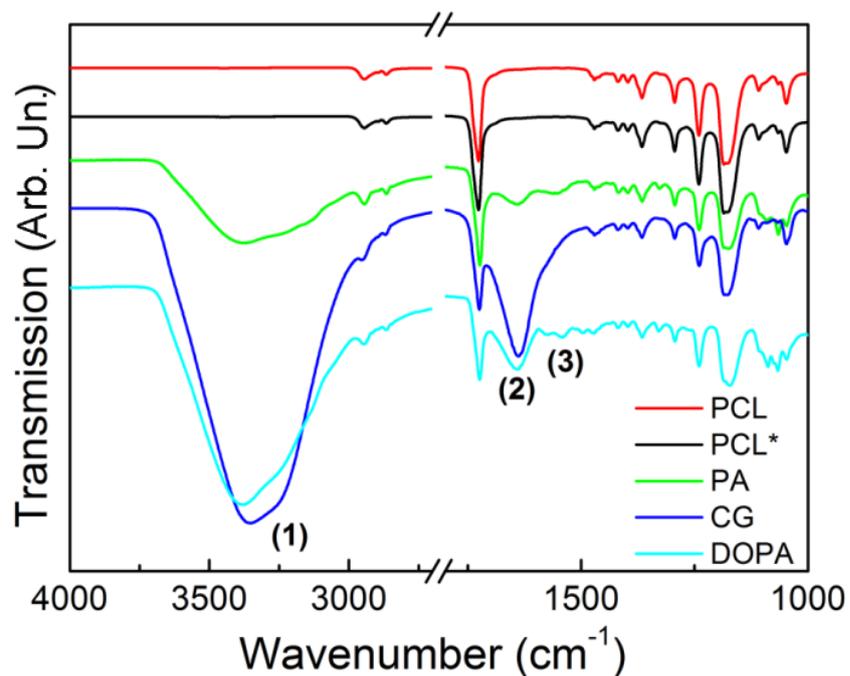

**Figure S2.** FTIR spectra of pristine PCL fibers (PCL, red curve), polydopamine-modified fibers without silicatein coating (PCL*, black curve) and silicatein-functionalized fibers by physisorption (*PA*, green curve), chemical grafting (*CG*, blue curve) and polydopamine (*DOPA*, cyan curve).

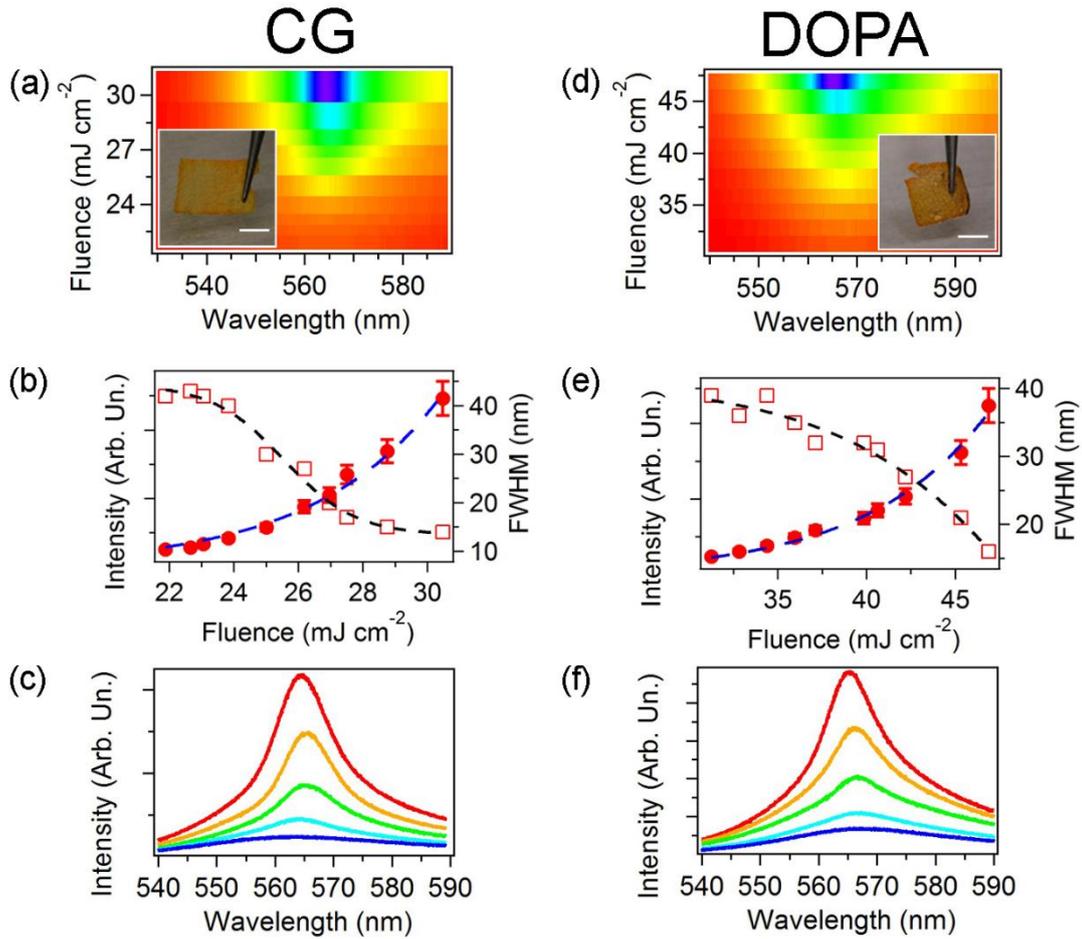

**Figure S3.** Analysis of the emission properties, under pulsed excitation, of *CG* (a-c) and *DOPA* dye-doped surfaces (d-f). (a), (d): Emission intensity maps showing the spectra of the samples as a function of the excitation fluence. Insets: corresponding device photographs. Scale bars = 5 mm. (b), (e): Intensity (left axis) and spectral linewidth (right axis) of the emission *vs.* excitation fluence. The blue dashed lines are fits to intensity data by an exponential-like law. The black dashed lines are guide to the eye for line-narrowing data. (c), (f): Emission spectra collected at increasing excitations fluences. Shown curves are averages of many single-shot emission spectra. From bottom to top, excitation fluences: (c): 23, 25, 27, 29, 30 mJ cm$^{-2}$; (g): 34, 37, 42, 45, 47 mJ cm$^{-2}$.

**Gain properties of pyrromethene in solid matrices**

The gain properties of pyrromethene in solid oligomer and polymer matrices under ns-pulsed excitation have been largely studied, using both frontal and transversal excitation, and variable stripe methods [S1-S3]. In two-dimensional slabs constituted by organic films deposited on quartz substrates, amplified spontaneous emission (ASE) is assisted by light waveguiding along the polymer layer, and the output intensity $I_L$ is related to the effective net gain, $G_{eff}$, and to the length, $L$, of the excitation stripe, as follows:

$$I_L = \frac{I_p A(\lambda)}{G_{eff}} \cdot [e^{G_{eff}(\lambda) \cdot L} - 1]. \tag{S1}$$

In this expression, $I_p$ is the non-amplified photoluminescence intensity and $A(\lambda)$ in turn depends on the spontaneous emission cross section. The effective gain, $G_{eff}$, relates to the stimulated emission gain by $G_{eff} = G_{se} - \gamma$, where $G_{se} = \sigma_{se} \cdot N_{exc}$, $\sigma_{se}$ is the stimulated emission cross-section and $N_{exc}$ indicates the excitation density.

The ASE spectrum of pyrromethene in solid matrices such as poly(methyl methacrylate) and polyvinylpyrrolidone shows a well-defined peak, centered at a wavelength between about 550 nm and 570 nm and with a spectral width of 10 nm. A maximum effective gain of about 20 cm$^{-1}$ has been measured at a pump intensity of 3.4 MW/cm$^2$ for pyrromethene-doped films [S1], leading to $\sigma_{se} \cong 5\text{-}7 \times 10^{-19}$ cm$^2$ (with $N_{exc} \sim 10^{19}$ cm$^{-3}$). By the value of the maximum optical gain, one can also extract the length, $l_{ASE}$, needed to find ASE, through the relation: $\exp(G_{eff} \cdot l_{ASE}) - 1 = 1$, i.e., $l_{ASE} = \ln 2 / G_{eff}$, which consequently is estimated to be 0.3 mm. While $G_{eff}$ has been obtained by exciting films by a frequency-doubled *Q*-switched Ng:YAG laser [S1], this value of $l_{ASE}$ is also in agreement with what we observe for pyrromethene/PVP films in our excitation scheme. We recall that the persistence of ASE components directionally guided along networks of even randomly-oriented electrospun nanofiber has been shown in previous work [S4].

The polymer matrix also helps in reducing oxidation screening the dyes and in improving photostability. This can be modelled considering oxygen diffusion through the polymer host towards the stationary, randomly-distributed chromophores acting as reactive traps which would result in fluorescence quenching. The intensity of the dye emission, which is proportional to the chromophore survival probability is then $I(t)/I_0 = \exp(-\gamma \rho^{\frac{2}{d+2}} t^{\frac{d}{d+2}})$, where $\gamma$ and $I_0$ are constants, $d$ is the spatial dimension of the diffusion medium, and $\rho$ is the density of quencher traps [S5]. For instance, for a simplified 1D case, this expression leads to [S6]:

$$\frac{I(t)}{I_0} = \exp\left[-\left(\frac{t}{\tau}\right)^{1/3}\right], \tag{S2}$$

where $\tau = \left[\left(\frac{3}{2}\right)^3 2\pi^2 \rho^2 D_{1D}\right]^{-1}$ is a characteristic time which depends on $n$ and on the diffusion coefficient, $D_{1D}$. In air atmosphere and ns-excitation, this is shown to lead to dye emission lifetimes of the order of $10^5$ pumping pulses [S3,S7,S8], further improving under vacuum.

**Coherent backscattering experiments**

The experimental set-up used for the measurement of the angular distribution of the coherent backscattered (CBS) intensity is based on a continuous wave He-Ne laser source with an emission wavelength of 633 nm. The laser output beam was expanded by a system of lenses and sent through a cube beamsplitter and a $\lambda/4$ waveplate before interacting with the sample. The incident beam had a diameter of 4 mm and typical power < 0.5 mW. The light backscattered by the sample was collected through the cube beamsplitter and a lens (with focal distance, *f*=30 cm), and filtered by a polarizer. The quarter-wave plate and the polarization filter were used to decrease contributions from single-scattering processes, which do not contribute to the enhanced backscattered cone, but increase the background signal and decrease the coherent backscattering enhancement factor. Indeed, while the CBS signal has helicity-conserving properties, single scattering processes do not preserve light helicity, thus allowing the contribution from single scattering events to be significantly suppressed by exploiting circularly polarized incident light and polarization filtering of the backscattered light [S9,S10]. The angular profile of the backscattered intensity was measured by a CCD detector (mod. iDus 40, Andor Technology), which was cooled down to -70 °C by Peltier elements. The typical integration time of a single acquisition was 0.5 s. Moreover, in order to minimize the presence of intensity peaks that are typically present in solid samples and might hide the CBS cone [S9], a mechanical system was used to displace periodically the sample by a few millimeters in a plane perpendicular to the propagation direction of the incident beam, at a typical frequency of 45 Hz. This experimental procedure allows the contribution from the speckle pattern to be significantly reduced (by about a factor 10) compared to the amplitude of the CBS peak [S9].

**Multiple scattering in finite slabs**

The fitting function of the CBS is defined as follows:

$$I(\theta) = [1 + (A_{CBS} - 1)(\gamma_C/\gamma_L)] \tag{S3}$$

where $A_{CBS}$ is the experimental enhancement factor, which deviates from its theoretical value 2 due to possible residual components from single or recurrent scattering, or for stray light. In the previous expression, the bistatic coefficients $\gamma_C$ (coherent contribution) and $\gamma_L$ (incoherent diffuse background) are defined as the measured scattered flux per solid angle and per unit of the probed area, normalized to the incident flux. They are obtained by the integral over all scattering paths [S11-S13]. The coherent intensity for a finite slab of thickness $L$ is shown in the following expression:

$$\gamma_C(\theta) = (3e^{-uL}) \times [2(-\alpha^2 + u^2 + \eta^2)\cosh(2\alpha z_0)\cos(L\eta) + 4\alpha\eta\sinh(2\alpha z_0)\sin(L\eta) + 2(-\alpha^2 - u^2 - \eta^2)\cos(L\eta) - 2(\alpha/u)(\alpha^2 - u^2 + \eta^2)\sinh(\alpha L + 2\alpha z_0)\sinh(uL) + 2(\alpha^2 - u^2 - \eta^2)\cosh(\alpha L + 2\alpha z_0)\cosh(uL) - 4\alpha u\sinh(\alpha L)\sinh(uL) + 2(\alpha^2 + u^2 + \eta^2)\cosh(\alpha L)\cosh(uL)] / \{-2\alpha\ell_t^3\sinh(\alpha L + 2\alpha z_0)[(-\alpha^2 + u^2 + \eta^2)^2 + (2\alpha\eta)^2]\} \tag{S4}$$

where $\theta$ is the scattering angle. The parameters entering in the function are so defined: $\eta \equiv \kappa(1-\mu_S)$, $u \equiv (1/2) \times \kappa_e(1 + \mu_S^{-1})$, and $\alpha \equiv (\ell_{abs}^{-2} + q_\perp^2)^{1/2}$ with $q_\perp = (2\pi/\lambda)\sin\theta$. Here $\mu_S = \cos\theta$, $z_0 = \alpha\ell_t$ is the extrapolation length that accounts for internal reflections at the boundaries ($z = 0$ and $z = L$). $\kappa_e = \ell_t^{-1} + \ell_i^{-1}$ is the extinction coefficient accounting for the attenuation of scattered intensity and $\alpha$ is the diffusive extinction coefficient, $\ell_t$ is the transport mean free path and $\ell_i$ is the inelastic (absorption) mean free path, while $\ell_{abs} = (\ell_t \ell_i/3)^{1/2}$.

The bistatic coefficient representing the diffuse background contribution is the following:

$$\gamma_L(\theta) = 3 \times [Z_1(1 + e^{-2L}) + Z_2(1 - e^{-2L}) + Z_3 e^{-L(u+\nu)}] / \{-2\alpha\ell_t^3\sinh(\alpha L + 2\alpha z_0) \times u[(u^2 - \alpha^2)^2 + \nu^2(\nu^2 - 2\alpha^2 - 2u^2)]\} \tag{S5}$$

where

$$Z_1 = u(\nu^2 - u^2 + \alpha^2)\cosh(\alpha L + 2\alpha z_0) + u(-\nu^2 + u^2 + \alpha^2)\cosh(\alpha L) - 2u\nu\alpha\sinh(\alpha L + 2\alpha z_0) - u\nu\alpha\sinh(\alpha L) \times (\nu^2 - 3u^2 - \alpha^2)/(u^2 - \alpha^2) \tag{S6a}$$

$$Z_2 = \nu(-\nu^2 + u^2 + \alpha^2)\cosh(\alpha L + 2\alpha z_0) - 2u^2\alpha\cosh(\alpha L) + \alpha(+\nu^2 + u^2 - \alpha^2)\sinh(\alpha L + 2\alpha z_0) + u^2\nu\cosh(\alpha L) \times (\nu^2 - u^2 - 3\alpha^2)/(u^2 - \alpha^2) \tag{S6b}$$

$$Z_3 = 2u(\nu^2 - u^2 - \alpha^2) + 2u(-\nu^2 + u^2 - \alpha^2)\cosh(2\alpha z_0) + 4u\nu\alpha\sinh(2\alpha z_0) \tag{S6c}$$

The function $\gamma_L$ is evaluated at $q_\perp = 0$, as a consequence $\alpha \equiv (\ell_{abs}^{-2})^{(1/2)}$; the parameter $\nu \equiv (1/2) \times \kappa_e(1 - \mu_S^{-1})$ is also introduced in the expression above.

**Light scattering by non-spherical particles in the T-matrix formalism**

In order to calculate the light scattering by non-spherical or composite particles we consider the particle embedded into a homogeneous, isotropic, indefinite medium (PVP in our case) of real refractive index $n$. The incident field is the polarized plane wave:

$$E_I = E_0\hat{e}_I exp(ik_I \cdot r), \tag{S7}$$

with unit polarization vector $\hat{\mathbf{e}}_I$ and propagation vector $\mathbf{k}_I = k\hat{\mathbf{k}}_I$, where $k=n k_v$ and $k_v=\omega/c$. The total field outside the particle is the sum of the incident and scattered field. The scattered field is obtained applying boundary conditions across the surface of the particle linking the internal and external fields. The overall scattering problem can be solved by the transition matrix (T-matrix) method [S14-S19], based on the definition of a linear operator relating the incident field to the scattered field [S14].

The starting point of the method is the field expansion in terms of the spherical multipole fields, i.e., the vector solutions of the Maxwell equations in a homogeneous medium that are simultaneously eigenfunctions of the angular momentum and the parity operators. The multipole expansion of the plane wave electric field and of the corresponding scattered wave is [S15]:

$$\mathbf{E}_I = E_0 \sum_{plm} \mathbf{J}_{lm}^{(p)}(\mathbf{r}, \mathbf{k}) W_{lm}^{(p)}(\hat{\mathbf{e}}_I, \hat{\mathbf{k}}_I) \tag{S8a}$$

$$\mathbf{E}_S = E_0 \sum_{plm} \mathbf{H}_{lm}^{(p)}(\mathbf{r}, \mathbf{k}) A_{lm}^{(p)}(\hat{\mathbf{e}}_I, \hat{\mathbf{k}}_I) \tag{S8b}$$

where $\mathbf{J}_{lm}^{(p)}(\mathbf{r}, \mathbf{k})$ denotes vector multipole fields that are regular at the origin of the coordinate system fixed inside the particle (Bessel multipoles), superscript $p$=1, 2 refers to the values of the parity index, distinguishing magnetic multipoles ($p$=1) from electric ones ($p$=2), the $\mathbf{H}_{lm}^{(p)}$ denotes multipole fields which satisfy the radiation condition at infinity (Hankel multipoles) and they are identical to the $\mathbf{J}_{lm}^{(p)}$ multipole fields except for the substitution of the Hankel function of the first kind $h_l(kr)$ in place of the spherical Bessel functions $j_l(kr)$.

The scattering process can be fully described by a linear operator $S$ defined by the equation:

$$\mathbf{E}_S = S\mathbf{E}_I \tag{S9}$$

The operator $S$ (transition operator) can be introduced because of the linearity of the Maxwell equations and of the equations expressing the boundary conditions across the surface of the particle. The $S$ representation on the basis of the spherical multipole fields gives the T-matrix defined by the equation [S14-S17]:

$$A_{lm}^{(p)} = \sum_{p'l'm'} S_{lml'm'}^{(pp')}(r, k) W_{l'm'}^{(p')}(\hat{\mathbf{e}}_I, \hat{\mathbf{k}}_I) \tag{S10}$$

relating the multipole amplitudes of the incident field, $W_{l'm'}^{(p')}(\hat{\mathbf{e}}_I, \hat{\mathbf{k}}_I)$, to those of the scattered field, $A_{lm}^{(p)}$. The elements of the T-matrix, $S_{lml'm'}^{(pp')}(r, k)$, contain all the information about the scattering process, but are independent of the state of polarization of the incident field. The transformation properties of the multipole fields under rotation of the coordinate frame imply corresponding transformation properties of the T-matrix under rotation of the scattering particle [S19,S20]. This is applicable to the calculation of the orientational average of all the quantities of interest for any choice of the orientational distribution function of the particles. Moreover, once the T-matrix has been calculated for an arbitrary orientation of one of the particles in an ensemble with respect to the incident field, it is straightforward to calculate all the scattering characteristics for any other particle orientation.

### Aggregates of spheres

The T-matrix approach solves the scattering problem by an aggregate (cluster) of spheres without resorting to any approximation [S21,S22]. The cluster model is very useful since it allows many cases of practical interest, dealing with nonspherical scatterers, to be simulated. Here we used it to model the scattering properties of fiber mats and silica-fibers hybrid structures.

We consider a cluster of spheres numbered by the index $\alpha$, of radius $\rho_\alpha$, and refractive index $n_\alpha$, whose centers lie at $\mathbf{R}_\alpha$. The spheres may have different refractive index and different radii. The field scattered by the whole aggregate is written as the superposition of the fields scattered by the single spheres:

$$\mathbf{E}_{S\eta} = E_0 \sum_\alpha \sum_{plm} \mathbf{H}_{lm}^{(p)}(r_\alpha, k) A_{\eta\alpha lm}^{(p)} \tag{S11}$$

where $\mathbf{r}_\alpha = \mathbf{r} - \mathbf{R}_\alpha$. The field within each sphere is taken in the form:

$$\mathbf{E}_{T\eta\alpha} = E_0 \sum_{plm} \mathbf{J}_{lm}^{(p)}(r_\alpha, k) C_{\eta\alpha lm}^{(p)} \tag{S12}$$

with $k_\alpha = n_\alpha k_\nu$, so that it is regular everywhere inside the sphere. The amplitudes $A_{\eta\alpha lm}^{(p)}$ and $C_{\eta\alpha lm}^{(p)}$ are determined by applying the boundary conditions across the surface of each of the spheres on the field. Now, the scattered field is given by a linear combination of multipole fields with different origins, whereas the incident field is given by a combination of multipole fields centered at the origin of the coordinates. Since the boundary conditions must be imposed at the surface of each of the spheres, e.g. of the $\alpha$–th sphere, we used the addition theorem [S15] to write the whole field in terms of multipole fields centered at $\mathbf{R}_\alpha$. In this way the scattered field at the surface of the $\alpha$–th sphere turns out to be [S15,S21]:

$$\mathbf{E}_{S\eta} = E_0 \left[ \sum_{plm} \mathbf{H}_{lm}^{(p)}(\mathbf{r}_\alpha, k) A_{\eta\alpha lm}^{(p)} + \sum_{\alpha'} \sum_{p'l'm'} \mathbf{J}_{lm}^{(p)}(r_\alpha, k) H_{\alpha lm\ \alpha'l'm'}^{(pp')} A_{\eta\alpha'l'm'}^{(p)} \right] \tag{S13}$$

where the operator $H$ is a transfer matrix. Analogously, the incident field at the surface of the $\alpha$–th sphere is [S15,S21]:

$$\mathbf{E}_{I\eta} = E_0 \sum_{plm} \sum_{p'l'm'} \mathbf{J}_{lm}^{(p)}(\mathbf{r}_\alpha, k) J_{\alpha lm\ 0l'm'}^{(pp')} W_{I\eta l'm'}^{(p')} \tag{S14}$$

where $\mathbf{R}_0 = 0$ is the vector position of the origin. Applying the boundary conditions, we get for each $\alpha$ four equations among which the elimination of the amplitudes of the internal field can be applied. Once this elimination is done, we get a system of linear inhomogeneous equations:

$$\sum_{\alpha'} \sum_{p'l'm'} M_{\alpha lm\ \alpha'l'm'}^{(pp')} A_{\eta\alpha lm}^{(p)} = -\mathcal{W}_{I\eta\alpha lm}^{(p)}, \tag{S15}$$

where we define:

$$\mathcal{W}_{I\eta\alpha lm}^{(p)} = \sum_{p'l'm'} J_{\alpha lm\ 0l'm'}^{(pp')} W_{I\eta\alpha lm}^{(p')} \tag{S16a}$$

and

$$M_{\alpha lm\ \alpha'l'm'}^{(pp')} = (R_{\alpha l}^{(p)})^{-1} \delta_{\alpha\alpha'} \delta_{pp'} \delta_{ll'} \delta_{mm'} + H_{\alpha lm\ \alpha'l'm'}^{(pp')} \tag{S16b}$$

The quantities $R_{\alpha l}^{(p)}$ are the Mie coefficients for the scattering from the $\alpha$–th sphere. The quantities $H_{\alpha lm\ \alpha'l'm'}^{(pp')}$ which come from the addition theorem, describe the multiple scattering processes occurring among the spheres in the aggregate. The occurrence of these processes is indicative of inter-spheres coupling. The elements $H_{\alpha lm\ \alpha'l'm'}^{(pp')}$ of the transfer matrix couple multipole fields both of the same and of different parity, with origin on different spheres. The formal solution to system (S15):

$$A_{\eta\alpha lm}^{(p)} = -\sum_{p'l'm'} [M^{-1}]_{\alpha lm\ \alpha'l'm'}^{(pp')} \mathcal{W}_{I\eta\alpha'l'm'}^{(p')} \tag{S17}$$

relates the multipole amplitudes of the incident field to those of the field scattered by the whole object. In order to define the T-matrix for the whole aggregate it is necessary to express the scattered field in terms of multipole fields with the same origin. Thanks to the addition theorem, the scattered field can be written as:

$$\mathbf{E}_{S\eta} = \left[ \sum_{plm} \sum_{\alpha'} \sum_{p'l'm'} \mathbf{H}_{lm}^{(p)}(r_\alpha, k) J_{0lm\ \alpha'l'm'}^{(pp')} A_{\eta\alpha'l'm'}^{(p')} \right] = \sum_{plm} \mathbf{H}_{lm}^{(p)}(\mathbf{r}, k) \mathcal{A}_{\eta\alpha lm}^{(p)} \tag{S18}$$

which is valid in the region external to the smallest sphere including the whole aggregate, and thus gives the scattered field in the far region. The previous equation shows that the field scattered by the whole cluster can be expanded as a series of vector multipole fields with a single origin (monocentered expansion) provided that the amplitudes are:

$$\mathcal{A}^{(p)}_{\eta\alpha lm} = \sum_{\alpha'p'l'm'} J^{(pp')}_{0lm\,\alpha'l'm'} A^{(p')}_{\eta\alpha'l'm'} \tag{S19}$$

Substituting for the amplitudes $\mathcal{A}$ their expression (S17), we can define the T-matrix of the aggregate as:

$$S^{(pp')}_{lml'm'} = -\sum_{\alpha\alpha'qq'll'mm'} J^{(pq)}_{0lm\alpha LM} [M^{-1}]^{(qq')}_{\alpha LM\,\alpha'L'M'} J^{(q'p')}_{\alpha L'M'\,0l'm'} \tag{S20}$$

In order to calculate the T-matrix of a cluster we have to solve the system (S15) which has, in principle, infinite order. Of course, the system must be truncated to some finite order by including into the multipole expansions terms up to order $l_M$, chosen to ensure the convergency of the calculations. For a cluster of $N$ spheres this implies solving a system of order $d_M = 2Nl_M(l_M + 2)$ which may grow rather big. Actually, the inversion of the matrix $M$ is responsible for most of the CPU time required for the calculation, which scales as $d_M^3$. As a consequence, the computational demand increases with the cube of the number of the spheres.

### *Light scattering angular dependence from nanofibers and hybrid structures*

To further investigate the light scattering of fiber mats and silica-fiber hybrid structures we calculated the scattered field intensity maps at 570 nm at different incident directions, $\hat{k}_{inc}$ (Figure S4). Normal incidence corresponds to $\hat{k}_{inc} = 0$ (Fig. S4a,b), whereas $\hat{k}_{inc} = 90$ corresponds to in-plane propagation (Fig. S4g,h). At all incident angles the hybridization of the nanofiber mats by the silica particles is crucial to increase the scattered fields.

Furthermore we studied how the scattered field intensity decays away from the individual particles out of the nanofiber mat plane (here, *x-y* is the plane of the fibers sample, and *z* is the direction perpendicular to this plane). An example of this out-of-plane decay and a typical map at 355 nm are shown in Figure S5a and S5b, respectively. These findings support that the stronger scattering is obtained for the silica hybridization at 355 nm. Moreover, the near-field decays by one order of magnitude within a distance of the order of the scatterer size (few hundreds of nanometers).

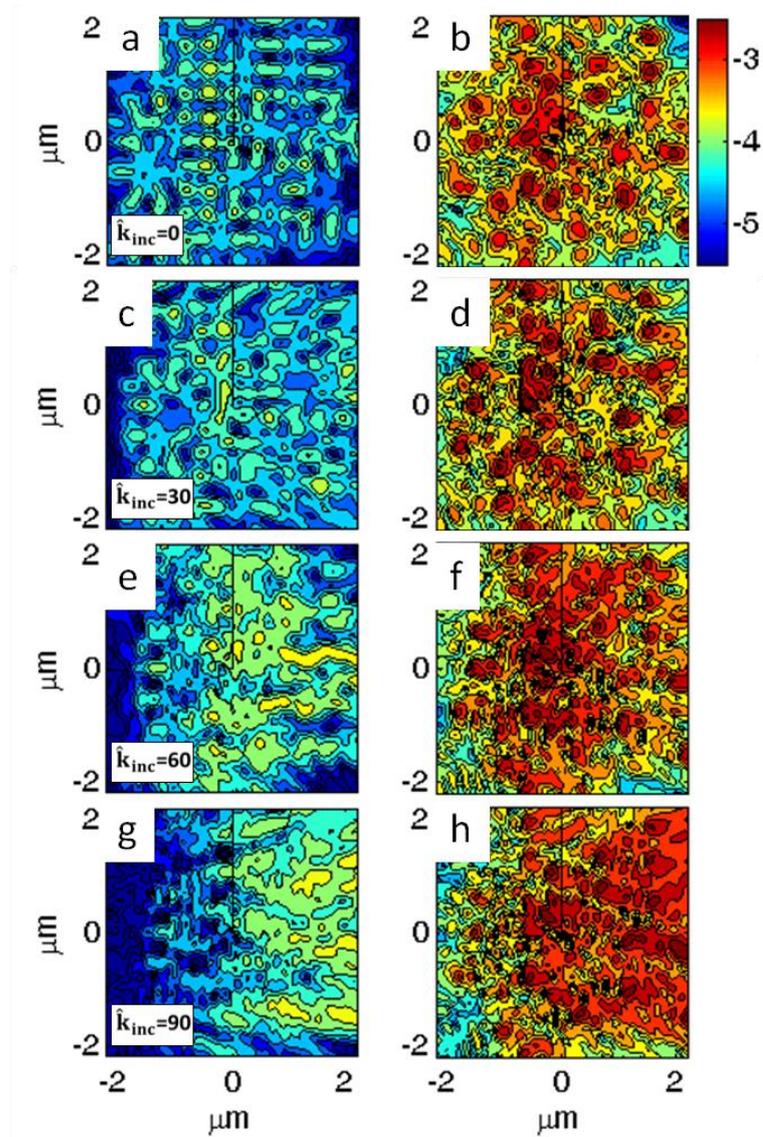

**Figure S4.** Normalized intensity maps, $|E_s/E_0|^2$, of the scattered field at 570 nm for the fiber mats (a, c, e, g) and for the hybridized structure embedding silica particles (b, d, f, h) at different incident angles with respect to the nanofiber mat plane, in logarithmic scale. $\hat{k}_{inc} = 0$ (a,b): Normal incidence. $\hat{k}_{inc} = 90$ (g,h): in-plane light propagation.

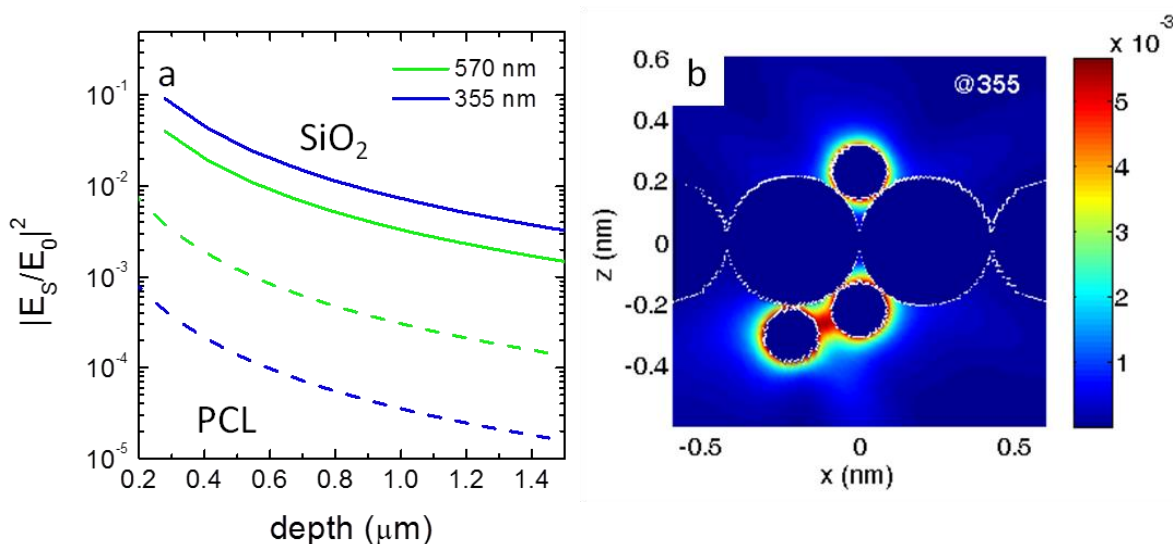

**Figure S5.** Normalized scattered field intensity decay in the out-of-plane ($z$) direction (depth). (a) Scattered field intensity vs. position from the particle centres for the PCL (dashed lines) and the silica (solid lines) particles, at different wavelengths. (b) Example of an intensity map in the $x$-$z$ plane at 355 nm. The small silica particles scatter the UV light much more than the larger PCL particles, supporting local field enhancement.

**References**

[S1] A. Costela, O. García, L. Cerdán, I. García-Moreno, and R. Sastre, Opt. Exp. **16**, 7023 (2008).

[S2] A. Costela, I. García-Moreno, L. Cerdán, V. Martin, O. García, and R. Sastre, Adv. Mater. **21**, 4163 (2009).

[S3] R.R. Valiev, E.N. Telminov, T.A. Solodova, E.N. Ponyavina, R.M. Gadirov, G.V. Mayer, and T.N. Kopylova, Chem. Phys. Lett. **588**, 184 (2013).

[S4] G. Morello, M. Moffa, S. Girardo, A. Camposeo, and D. Pisignano, Adv. Function. Mater. **24**, 5225 (2014).

[S5] S. Redner and K. Kang, J. Phys. A: Math. Gen. **17**, L451 (1984).

[S6] G.S. Kanner, X. Wei, B.C. Hess, L.R. Chen, and Z.V. Vardeny, Phys. Rev. Lett. **69**, 538 (1992).

[S7] A. Camposeo, F. Di Benedetto, R. Stabile, R. Cingolani, and D. Pisignano, Appl. Phys. Lett. **90**, 143115 (2007).

[S8] A. Costela, I. García-Moreno, J.M. Figuera, F. Amat-Guerri, R. Mallavia, M.D. Santa-Maria, and R. Sastre, J. Appl. Phys. **80**, 3167 (1996).

[S9] S.R. Etemad, R. Thompson, and M.J. Andrejco, Phys. Rev. Lett. **57**, 575 (1986).

[S10] O.L. Muskens and A. Lagendijk, Opt. Exp. **16**, 1222 (2008).

[S11] M.B. van der Mark, M.P. van Albada, and A. Lagendijk, Phys. Rev. B **37**, 3575 (1988).

[S12] E. Akkermans, P.E. Wolf, R. Maynard, and G. Maret, J. Phys. France **49**, 77 (1988).

[S13] D.S. Wiersma, PhD thesis. University of Amsterdam (1995).

[S14] P.C. Waterman, Phys. Rev. D **3**, 825 (1971).

[S15] F. Borghese, P. Denti, and R. Saija, Scattering from model nonspherical particles, 2$^{nd}$ Ed., Springer-Verlag Berlin Heidelberg (2007).

[S16] F. Borghese, P. Denti, R. Saija, M.A. Iatì, and O.I. Sindoni, J. Quant. Spectrosc. Radiat. Transfer **70**, 237 (2001).

[S17] M.A. Iatì, C. Cecchi-Pestellini, D.A. Williams, F. Borghese, P. Denti, R. Saija, and S. Aiello, Monthly Notices of the Royal Astronomical Society **322**, 749 (2001).

[S18] M.I. Mishchenko, L.D. Travis, and A.A. Lacis, Scattering, absorption, and emission of light by small particles, Cambridge University Press (2002).

[S19] R. Saija, M.A. Iatì, F. Borghese, P. Denti, S. Aiello, and C. Cecchi-Pestellini, Astrophys. J. **559**, 993 (2001).

[S20] N.G. Khlebtsov, Appl. Opt. **31**, 5359 (1992).

[S21] F. Borghese, P. Denti, R. Saija, G. Toscano, and O.I. Sindoni, Aerosol Sci. Technol. **3**, 227 (1984).

[S22] F. Borghese, P. Denti, R. Saija, G. Toscano, and O.I. Sindoni, Il Nuovo Cimento B (1971-1996) **81**, 29 (1984).